\documentclass{aa}  

\usepackage{graphicx}
\usepackage{xcolor}
\usepackage{txfonts}
\begin{document} 

%DA CAMBIARE
%   \title{Binaries among split main sequence stars to constrain the formation}
% APM20apr: metterei il paper all'interno della serie. 
\title{{\it Hubble Space Telescope} survey of Magellanic Cloud star clusters. 
 Binaries among the split main sequences of NGC\,1818, NGC\,1850, and NGC\,2164.}

   \author{F. Muratore \inst{1}, %\fnmsep
        A. P. Milone \inst{1,2}, %\fnmsep
        F. D'Antona \inst{3}, %\fnmsep
        E. J. Nastasio \inst{1},
        G. Cordoni \inst{4}, %\fnmsep
        M. V. Legnardi \inst{1}, %\fnmsep
        C. He \inst{5,6}, %\fnmsep
        T. Ziliotto \inst{1}, %\fnmsep
        E. Dondoglio \inst{2}, %\fnmsep
        M. Bernizzoni \inst{1}, %\fnmsep
        M. Tailo \inst{2}, %\fnmsep
        E. Bortolan \inst{1}, %\fnmsep
        F. Dell'Agli \inst{3}, %\fnmsep
        L. Deng \inst{7,8,9}, %\fnmsep
        E. P. Lagioia \inst{10}, %\fnmsep
        C. Li, \inst{5,6}, %\fnmsep
        A. F. Marino \inst{2}, %\fnmsep
        P. Ventura \inst{3} }%\fnmsep}

   \institute{Dipartimento di Fisica e Astronomia “Galileo Galilei”, Università Degli Studi di Padova, Vicolo dell’Osservatorio 3, 35122 Padova, Italia
     \and
     Istituto Nazionale di Astrofisica - Osservatorio Astronomico di Padova, Vicolo dell’Osservatorio 5, 35122 Padova, Italy 
     \and 
     Istituto Nazionale di Astrofisica, Osservatorio Astronomico di Roma, Via Frascati 33, 00077 Monte Porzio Catone, Italy
     \and
     Research School of Astronomy and Astrophysics, Australian National University, Canberra, ACT 2611, Australia  
      \and
      School of Physics and Astronomy, Sun Yat-sen University, Zhuhai 519082, China
      \and
      CSST Science Center for the Guangdong–Hong Kong-Macau Greater Bay Area, Zhuhai, 519082, People’s Republic of China
      \and
      Key Laboratory for Optical Astronomy, National Astronomical Observatories, Chinese Academy of Sciences, Beijing 100101, People’s Republic of China
      \and
      School of Astronomy and Space Science, University of Chinese Academy of Sciences, Beijing 100049, People’s Republic of China
      \and
      Department of Astronomy, School of Physics and Astronomy, China West Normal University, Nanchong 637002, People’s Republic of China
      \and
      South-Western Institute for Astronomy Research, Yunnan University, Kunming 650500, PR China 
%      \and
%     Istituto Nazionale di Astrofisica - Osservatorio Astrofisico di Arcetri, Largo Enrico Fermi, 5, I-50125 Firenze, Italy
      \\    
}

   \date{Received May XX, XXXX; accepted May XX, XXXX}
\titlerunning{Binaries among split main sequences in young star clusters} 
\authorrunning{Muratore et al.}

% \abstract{}{}{}{}{} 
% 5 {} token are mandatory

\abstract
{Nearly all star clusters younger than $\sim$600 Myr exhibit extended main sequence turn offs and split main sequences (MSs) in their color-magnitude diagrams. Works based on both photometry and spectroscopy have firmly demonstrated that the red MS is composed of fast-rotating stars, whereas blue MS stars are slow rotators. 
Nevertheless, the mechanism responsible for the formation of stellar populations with varying rotation rates remains a topic of debate. Potential mechanisms proposed for the split MS include binary interactions, early evolution of pre-main sequence stars, and the merging of binary systems, but a general consensus has yet to be reached. These formation scenarios predict different fractions of binaries among blue- and red-MS stars. Therefore, studying the binary populations can provide valuable constraints that may help clarify the origins of the split MSs.
We use high-precision photometry from the {\it Hubble Space Telescope} ({\it HST}) to study the binaries of three young Magellanic star clusters exhibiting split MS, namely NGC\,1818, NGC\,1850, and NGC\,2164. 
By analyzing the photometry in the F225W, F275W, F336W, and F814W filters for observed binaries and comparing it to a large sample of simulated binaries, we determine the fractions of binaries within the red and the blue MS.
We find that the fractions of binaries among the blue MS are higher than those of red-MS stars by a factor of $\sim 1.5$, $4.6$, and $\sim 1.9$ for NGC\,1818, NGC\,1850, and NGC\,2164, respectively. We discuss these results in the context of the formation scenarios of the split MS. }


   \maketitle
%
%-------------------------------------------------------------------

\section{Introduction}

High-precision photometry from the \textit{Hubble Space Telescope} (HST) and ground-based observatories have shown that Magellanic Cloud star clusters younger than approximately two billion years display complex color-magnitude diagrams (CMDs) that cannot be explained by simple isochrones \citep[see][for recent reviews]{milone2022, li2024}. The CMDs of nearly all clusters younger than 600 Myr exhibit extended main-sequence turn-offs (eMSTOs) and split main sequences \citep[MSs,][]{milone2015a, milone2018, dantona2015}, with the reddest sequence being the most populated.  In contrast, the star clusters with ages between $\sim$600 Myr and $\sim$2 Gyr show only the eMSTO \citep[e.g.][]{bertelli2003, mackey2007, glatt2008, milone2009, goudfrooij2011, correnti2014}.

Nowadays, it is widely accepted that stellar rotation plays a major role in shaping the split MSs and eMSTOs of young and intermediate-age star clusters \citep{bastian2009, dantona2015, niederhofer2015, cordoni2018, georgy2019}.
However, in addition to the rotation, factors such as age spread and dust have also been proposed as potential contributors to the observed eMSTOs \citep[see][for discussion]{goudfrooij2017, milone2017b, li2017, cordoni2022, dantona2023}.

The MSs of clusters younger than $\sim$600 Myr host rapidly rotating stars, which compose the red MS (rMS), and a slowly rotating stellar population for the blue MS (bMS). 
This evidence, first suggested by the comparison between isochrones with different rotation rates and the observed CMDs of clusters with split MSs \citep{dantona2015, milone2016}, has been proved by works based on high-resolution spectra of MS stars \citep{marino2018b, marino2018a}. Similarly, the eMSTO hosts stars with different rotation rates, with fast rotators populating the upper eMSTO \citep{dupree2017, bastian2018b, cristofori2024, cordoni2024}.

The physical reasons behind the presence of both fast-rotating stars and slow-rotators in Magellanic Cloud clusters remain unsettled.
Observations of field stars in the last century showed that indeed also field stars display a bimodal distribution, especially in the mass range between $2.4$ and $3.85$ $M_{\odot}$ \citep{zorec2012},
with a slow and a fast velocity component. A similar bimodality in the velocity distribution has also been detected among early B-type stars by \cite{dufton2013}. \cite{zorec2012} suggested two possible reasons for this velocity difference:
\begin{enumerate}
    \item The first possibility is based on the fact that it is unlikely that pre-MS stars originate with zero angular momentum. It is possible, however, that a fraction of them could have undergone magnetic braking;
    \item As another possibility, some form of tidal braking is occurring due to the presence of a stellar companion. This tidal braking may be significantly more efficient than that predicted by the dynamical tidal model proposed by \cite{zahn1975, zahn1977}. See \citet{dantona2015} and \citet{He2023} for discussions.
\end{enumerate}
These two hypotheses were revived more recently, the first one mainly by \cite{bastian2020}, and the second one by \cite{dantona2015,dantona2017}.
More specifically, stars may not be able to accelerate during the pre-MS contraction if they are magnetically locked with the accretion disk, and they can therefore reach the MS with reduced rotation. As a consequence, the presence of bMS and rMS stars in young star clusters would depend on the early evolution of pre-MS stars, particularly on whether they retain or lose their protoplanetary discs during the first few million years.

The alternative is that interactions in binary star systems are responsible for stellar braking. In this case, tidal interactions would lead to braking effects within the core, extending outward. Stars would only reach their final position on the non-rotating sequence once the braking process extends to the stellar envelope and photosphere. If this occurs, the star would shift to the bMS.
\cite{dantona2015} adopted this view to interpret the split MS of the Large Magellanic Cloud (LMC) star cluster NGC\,1856, quoting the study of A- and B-type binary stars \citep{abt2004} in the Galactic field, where these authors find that close binaries with periods between 4 and 500 days rotate significantly slower than single stars.
This process would lead to a higher binary fraction in slow rotators than in the rapidly rotating population, and the bMS would predominantly consist of binaries with small mass ratios.

\cite{wang2022} suggested that binary mergers of MS stars could be responsible for generating slow rotators. When the two stars in a binary merge, the outcome is a single star that has a core hydrogen content higher than that of a single star of the same mass and age. As a result, these merger products, although having the same age as other cluster stars, appear younger and exhibit a bluer color in the CMD. Possibly, in this case, we could expect to observe a lack of binaries among slow-rotating stars.

Studying binary systems among slow- and fast-rotating stars in clusters with split MS would help constrain the scenarios for forming stellar populations with different rotation rates. Pioneering work by \cite{kamann2021}, based on radial velocities and rotational velocities of stars in the LMC cluster NGC 1850, results in a comparable fraction of binaries among bMS and rMS stars, of $5.9\pm1.1\%$ and $4.5\pm0.6\%$, respectively.

In addition to providing insights into the origins of split MSs and eMSTOs, understanding the binary populations in star clusters is crucial for numerous astrophysical studies. Binaries significantly influence the dynamical evolution of clusters, serving as an important source of heating. Accurate measurements of the binary fraction are necessary for determining the stellar mass and luminosity functions in GCs. Moreover, stellar evolution in binary systems can differ from that of isolated stars, indeed it could lead to the formation of unique objects, such as blue stragglers, cataclysmic variables, millisecond pulsars, and low-mass X-ray binaries.

In this work, we use multi-band {\it HST} photometry obtained as part of our recent survey of Magellanic-Cloud stars clusters \citep{milone2023} to study the binaries in the young LMC clusters NGC\,1818 ($\sim$40 Myr), NGC\,1850 ($\sim$90 Myr), and NGC\,2164 ($\sim$100 Myr) and infer the incidence of binaries among bMS and rMS stars.

The paper is organized as follows. Section\,\ref{sec:data} describes the dataset, and Section\,\ref{sec:analysis} illustrates the methods we used to estimate the fraction of binaries among the bMS and the rMS. Finally, Section\,\ref{sec:summary} is dedicated to a summary and a discussion of the results.

\section{Data} \label{sec:data}
The data used in this work consist of images of NGC\,1818, NGC\,1850, and NGC\,2164 collected with the Ultraviolet and Visual Channel of the Wide Field Camera 3 (UVIS/WFC3) on board {\it HST}.
Stellar positions and photometry are derived by \cite{milone2023, milone2023b} as part of their {\it HST} survey of Magellanic Cloud star clusters. For NGC\,1818 and NGC\,2164 we used photometry in the F225W, F336W, and F814W bands of UVIS/WFC3, whereas for NGC\,1850 we exploited F275W, F336W, and F814W stellar magnitudes.

The data reduction has been carried out with the KS2, a computer program developed by Jay Anderson. Originally designed to reduce images collected with the Wide Field Channel of the Advanced Camera for Surveys onboard the {\it HST}, KS2 has evolved from the earlier “kitchen sync” program \citep{anderson2008}. 
 
In a nutshell, KS2 employs three distinct methods to measure stars, with each method optimizing astrometry and photometry for stars of varying luminosities. Since our focus is on bright stars, we utilized the results obtained from Method I.
This method takes advantage of the best effective point spread function \citep[ePSF,][]{anderson2000} model to measure all stellar sources that produce a distinct peak within a 5$\times$5 pixel region, following the subtraction of neighboring stars.
It calculates the flux and position of each star in each image separately, by using the ePSF model associated with its position. The sky level is subtracted considering an annulus between four and eight pixels from the stellar center. Finally, the results from all images are averaged to determine the best stellar magnitudes and positions. The photometry has been calibrated to the Vega-mag reference frame by using the zero points provided by the Space Telescope Science Institute website.
  
The KS2 program offers various diagnostics of photometric quality. In our analysis, we focused on isolated stars that are well-fitted by the ePSF model. For NGC\,1850 we used the stellar proper motions computed by \citet{milone2023b} by comparing the stellar positions measured in multi-epoch images. These proper motions are used to separate the bulk of field stars from cluster members. 

The photometry of NGC\,1850 has been corrected for the effects of differential reddening by using the reddening maps derived by \citet{milone2023b}. We verified that NGC\,1818 and NGC\,2173 are not significantly affected by differential reddening by adopting the criteria by \cite{legnardi2023}. Therefore, for these clusters, we used the original photometry.
  
We refer to section 2.4 in \cite{milone2023}, for further details on the images, the methods for measuring stellar positions and fluxes, and the criteria adopted to select the stars with high-precision photometry and astrometry.

\subsection*{Artificial star tests}
For each cluster, we conducted artificial star (AS) tests to estimate photometric errors and generate the simulated CMDs. Following the method outlined by \cite{anderson2008}, we created a list of 20,000 ASs. These ASs were designed to mimic the radial distributions and luminosity functions of the observed stars.
The instrumental magnitudes of the ASs spanned from approximately $-$13.6 mag to $-$9.0 mag in the UVIS/WFC3 F814W filter. For other filters, we derived magnitudes based on the fiducial lines of the rMS and the bMS.
To determine the magnitudes and positions of the ASs, we employed the KS2 program, using the same procedure that \cite{milone2023} applied to real stars. Our investigation focused exclusively on relatively isolated ASs that exhibited good fits to the PSF, and that were selected by following the same criteria as those used for the real stars.

\begin{figure*}
    \centering
    \includegraphics[width=1.\linewidth]{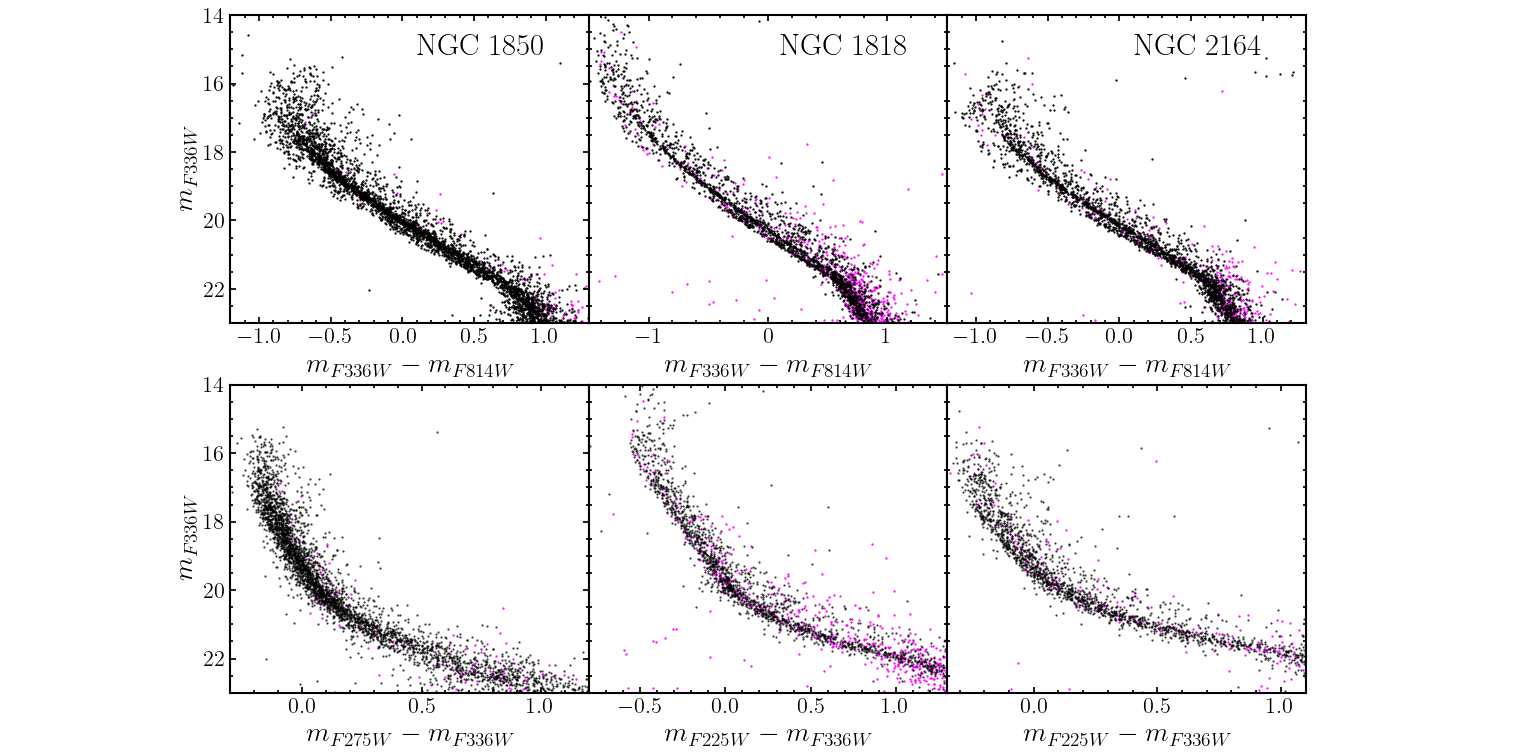}
    \caption{ A collection of CMDs of the studied clusters, namely NGC\,1850, NGC\,1818, and NGC\,2164. The stars in the cluster field are represented with black dots, whereas the violet dots refer to the reference-field stars. For NGC\,1850 we used stellar proper motions to identify and exclude from the analysis the bulk of field stars.}
    \label{fig:cmds}
\end{figure*}

In addition to the cluster members, the field of view observed by {\it HST} comprises field stars that lie along the same line of sight of each cluster.
To minimize the contamination from background and foreground stars, we restricted our analysis to a central field of view, hereafter cluster field. The latter is derived by using the procedure by \cite[][see their section 3.1]{mohandasan2024}. 
In summary, we estimated the number density of stars brighter than $m_{F336W}=22.0$ mag at different radial distances from the cluster center, and identified the radius where the density profile is nearly flat.
The portion of the {\it HST} field of view outside this radius is dominated by field stars and is named field region.
We assumed that the number density of field stars corresponds to the average stellar density in the field region. 
We defined the radius of the cluster region as the distance at which the observed stellar density is three times higher than the field density. 
The radii of the cluster and field regions, within we selected cluster members and field stars, are listed in Table\,\ref{tab1} for each cluster, whereas the $m_{\rm F336W}$ vs\,$m_{\rm F336W}-m_{\rm F814W}$ and $m_{\rm F336W}$ vs\,$m_{\rm F275W}-m_{\rm F336W}$ CMDs for stars in the cluster and reference fields are shown in Fig.\,\ref{fig:cmds}.

In the case of NGC\,1850 we took advantage of stellar proper motions to further minimize the contamination from field stars. Moreover, the proper motions allowed us to identify and exclude from the analysis the bulk of stars in BRHT\,5b, which is a young star cluster that lies on the same line of sight as NGC\,1850 \cite[see][for details on the proper motions of stars in the field of view of NGC\,1850]{milone2023b}.
In addition, we excluded stars on the west side of NGC\,1850 which includes the stellar overdensity named NGC\,1850A, which is often considered a separate cluster. 

\section{The incidence of binaries among red-MS and blue-MS stars}\label{sec:analysis}

In this section, we present the procedure used to derive the fractions of binary systems composed of blue MS and red MS stars (hereafter bMS and rMS binaries). 
Our method, based on the approach introduced by \citet{milone2020} to infer the fraction of binaries stars among the multiple populations of five Galactic globular clusters, consists of three main steps, which we will discuss in the following subsections.

Specifically, in Section\,\ref{sub:BinMSs} we estimate the fraction of bMS and rMS binaries with respect to a sample of binaries with similar luminosity ratios. Section\,\ref{sub:MSs} details the procedure used to derive the fractions of bMS, rMS, and binary stars. Finally, we combine these results in Section\,\ref{sub:results} to determine the fractions of bMS binaries relative to the total number of bMS stars and rMS binaries relative to the total number of rMS stars.

\begin{table}
    \centering
    \caption{Radii of the cluster field and minimum and maximum radii for the reference fields.}
    \begin{tabular}{ccc}
    \hline
    \hline
         Target& $r_{cluster}$ [arcsec]&  $r_{field}$ [arcsec]\\
         \hline
         NGC\,1818&   $32$ & $56-72$  \\
         NGC\,1850&   $32$ & $50-113$   \\ 
         NGC\,2164&   $32$ & $56-72$  \\
         \hline
    \end{tabular}
   % \vspace{0.2cm}
    \label{tab1}
\end{table}

\subsection{The fractions of red-MS and blue-MS stars among binaries} \label{sub:BinMSs}

\begin{figure*}
    \centering
    \includegraphics[width=1.\linewidth]{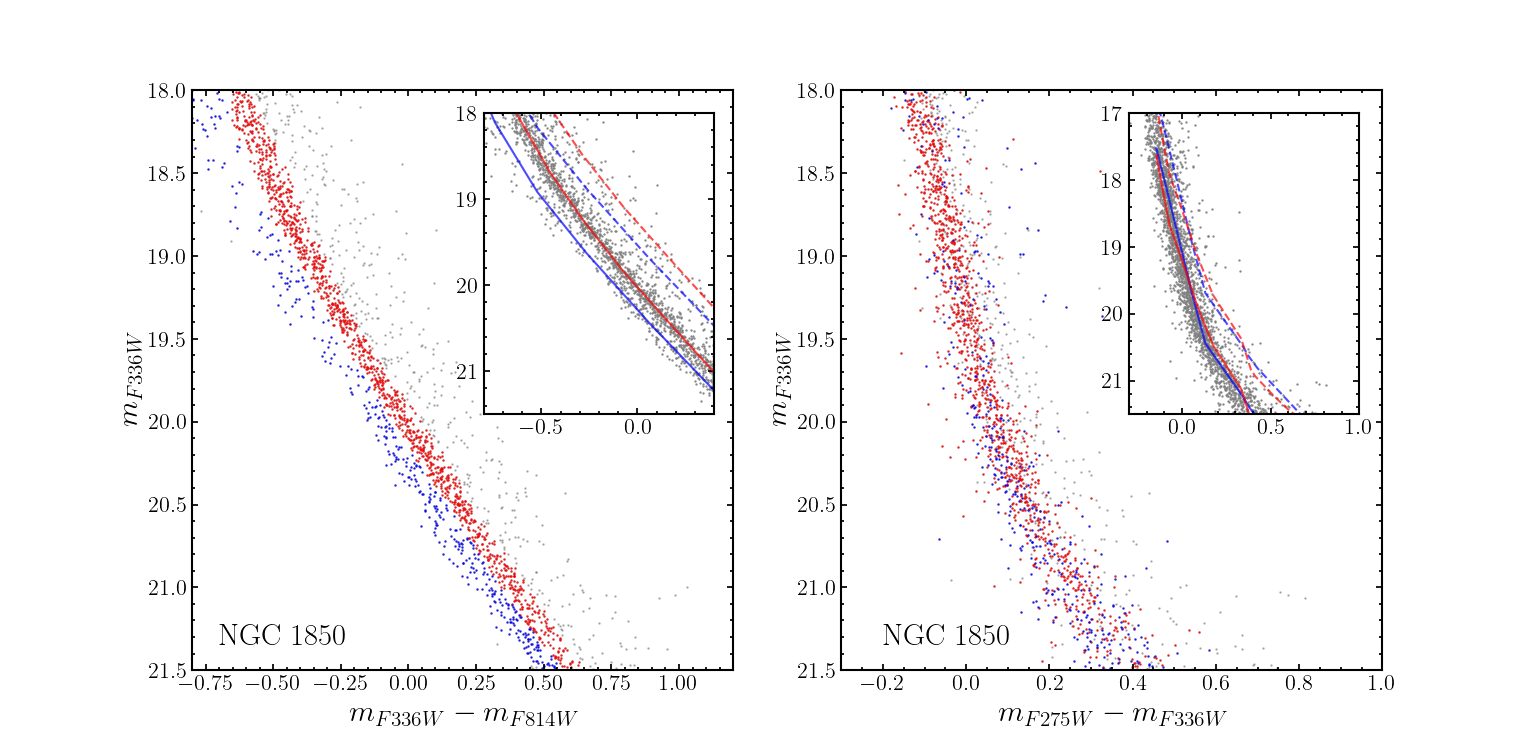}
    \caption{$m_{\rm F336W}$ vs.\,$m_{\rm F336W}-m_{\rm F814W}$ (left) and $m_{\rm F336W}$ vs.\,$m_{\rm F275W}-m_{\rm F336W}$ (right) CMDs of stars in the NGC\,1850 cluster field. The CMDs are zoomed-in around the luminosity interval that we used to investigate the incidence of binaries among the stellar populations with different rotation rates. The probable rMS and bMS stars, selected in the left-panel CMD, are colored red and blue, respectively, whereas the remaining stars are represented with gray dots. The insets show the fiducial lines of the red and blue MS (continuous lines), and the fiducial lines of equal-luminosity bMS and rMS binaries (dashed lines).}
    \label{ngc1850_fids}
\end{figure*}

\begin{figure*}
    \centering
    \includegraphics[width=1.\linewidth]{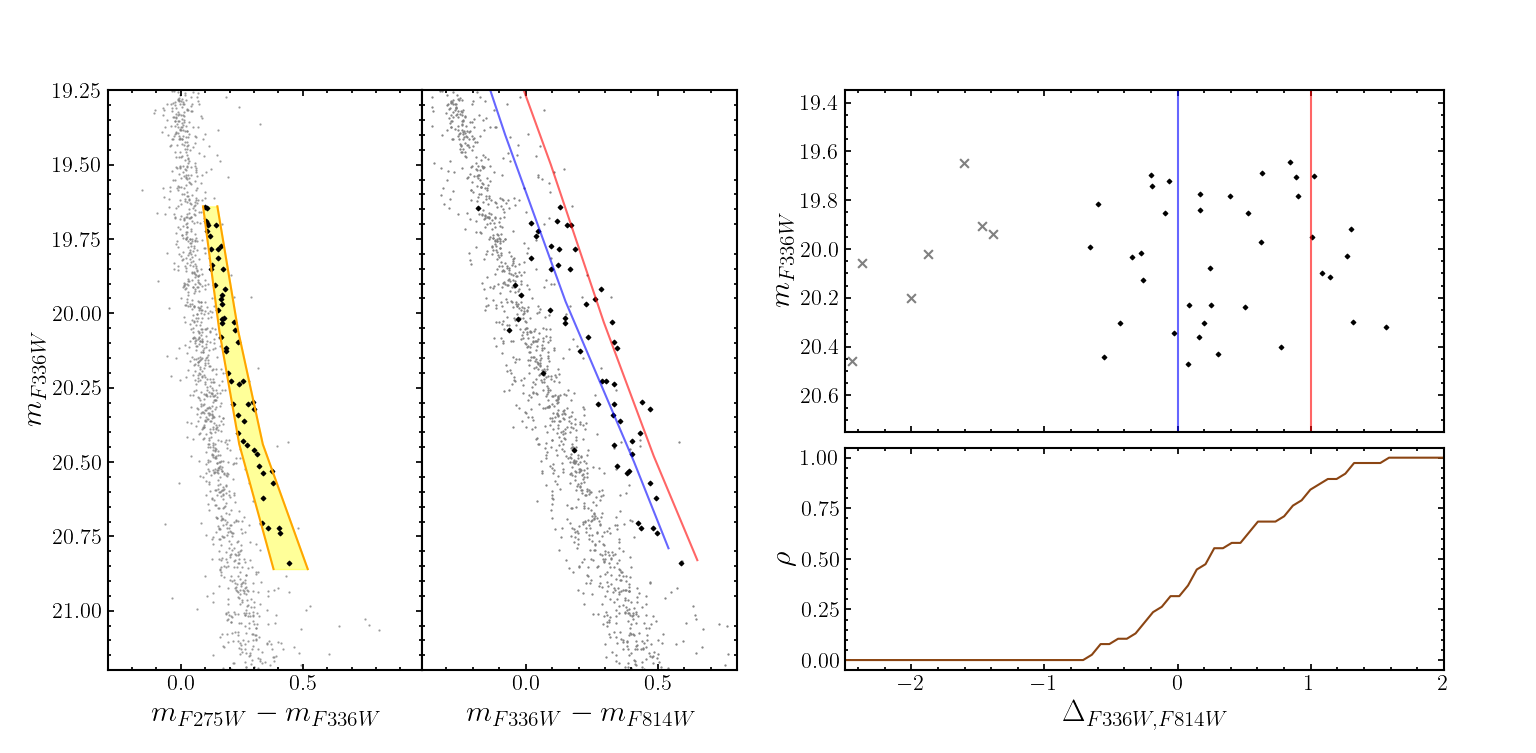}
    \caption{
    $m_{\rm F336W}$ vs.\,$m_{\rm F275W}-m_{\rm F336W}$ (left panel), $m_{\rm F336W}$ vs.\,$m_{\rm F336W}-m_{\rm F814W}$ CMDs (middle panel), the verticalized $m_{\rm F336W}$ vs.\,$\Delta_{\rm F336W,F814W}$
     diagram (top-right panel), and the $\Delta_{\rm F336W,F814W}$ cumulative distributions (bottom-right panel). The investigated binaries, selected in the left CMD and located within the shadow yellow region are colored black, whereas the remaining stars are represented with gray dots. 
    The gray crosses in the top-right panel are stars excluded from the analysis. The red and blue lines are the fiducial lines for equal-luminosity rMS and bMS binaries, respectively.}
    
    \label{ngc1850_oss}
\end{figure*}

Due to the large distance of the LMC, the binaries analyzed in this work are unresolved stellar systems that appear as point-like sources in the CMD. The magnitude of each system is:
\begin{equation}
%    m_{bin}=-2.5\log(F_1+F_2)= m_1-2.5\log(1+\frac{F_1}{F_2})
 m_{bin}=m_1-2.5\log{ \Biggl( 1+\frac{F_1}{F_2} \Biggl)}
\end{equation}

where $m_{1}$ is the magnitude of the brightest star and $F_1$ and $F_2$ are the fluxes of the primary and secondary components, respectively.
In a simple-population star cluster, a binary system composed of two MS stars with the same luminosity will result in a source $\sim$0.75 mag brighter than each component and the same color.
Hence, the equal-luminosity binaries populate a sequence that runs parallel to the MS, but $\sim$0.75 mag brighter.
Binaries composed of two MS stars with different luminosities are located in the CMD region between the MS and the equal-luminosity binaries sequences and their colors and magnitudes depend on the luminosities of the two components \citep[see][and references therein for details]{milone2012, mohandasan2024}.
The procedure that we used to estimate the fraction of binaries among bMS and rMS stars is illustrated in Figs.\,\ref{ngc1850_fids} and \ref{ngc1850_oss} for NGC\,1850 taken as a test case.
The method takes advantage of the $m_{\rm F336W}$ vs.\,$m_{\rm F336W}-m_{\rm F814W}$ CMD, wherein bMS and rMS stars delineate separate sequences. Additionally, it leverages the $m_{\rm F336W}$ vs.\,$m_{\rm F275W}-m_{\rm F336W}$ CMD, where the two sequences exhibit considerable overlap.

These CMDs are plotted in Fig.\,\ref{ngc1850_fids}, where we use red and blue colors to represent the sample of bona-fide bMS and rMS stars selected by eye from the $m_{\rm F336W}$ vs.\,$m_{\rm F336W}-m_{\rm F814W}$ CMD.

The insets of Fig.\,\ref{ngc1850_fids} show the fiducial lines of the red and blue MSs in solid and the fiducial lines of the equal-luminosity binary sequences in dashed. 
The solid lines are derived by linearly interpolating the median colors and magnitudes of the selected samples of stars over 0.5-wide F336W bins.
The dashed ones are evaluated by subtracting $\sim$0.75 mag from the magnitudes of the solid lines.

Our investigation relies on binary systems composed of stars with similar luminosities. As shown in the left panel of Fig.\,\ref{ngc1850_oss}, these binaries are selected from the $m_{\rm F336W}$ vs.\,$m_{\rm F275W} -m_{\rm F336W}$ CMD and are situated within the shaded yellow region.
Its boundaries are derived as follows. The bright boundary is set at $m_{F336W} = 19.6$ mag, as for brighter magnitudes binary stars are mixed with MS single stars. The faint boundary is set at $m_{F336W} = 20.8$ mag because it is challenging to disentangle blue and red MS binaries with fainter magnitudes.
 
The orange lines mark the blue and red limits of the shaded yellow region. They are obtained by adding and subtracting the corresponding color uncertainties to the colors of the equal-luminosity binaries fiducial line.
The selected sample comprises 55 candidate binary systems. The selected binaries are marked with black symbols in the $m_{\rm F336W}$ vs.\,$m_{\rm F336W} - m_{\rm F814W}$ CMD shown in the middle panel of Fig.\,\ref{ngc1850_oss}, where the bulk of selected binaries are located near the fiducial lines of equal-luminosity bMS and rMS binaries.

We derived the verticalized $m_{F336W}$ vs. $\Delta_{F336W, F814W}$ diagram, in such a way that the fiducial lines of equal-luminosity binaries result in vertical lines. Specifically, the equal-luminosity blue MS fiducial line is set at zero on the abscissa and the one referred to the red MS at one (as illustrated in the top-right panel of Fig.\,\ref{ngc1850_oss}). 

To minimize the contamination from field stars and stars with large photometric errors (including single stars and binaries with small luminosity ratio), we excluded the outliers with $\Delta_{\rm F336W, F814W} < -0.7$ mag and $\Delta_{\rm F336W, F814W} > 1.7$ mag.

The selected binaries are used to derive the cumulative distribution shown in the bottom right panel in Fig.\,\ref{ngc1850_oss}.

To infer the fraction of bMS and rMS binaries, we compared the  $\Delta_{\rm F336W,F814W}$ cumulative distribution of the observed binaries, with the corresponding distribution that results from simulated diagrams made with ASs.
To do this, we generated a grid of diagrams with different fractions of bMS ($f_{bMS,bin}$ ) and rMS ($f_{rMS,bin}$ ) binaries. 
In particular, we generated binaries from a flat distribution of mass ratio and then we selected those within the same yellow region used for real stars.
We assumed that $f_{rMS,bin}$  ranges from 0.00 to 1.00 in steps of 0.01 and  $f_{bMS,bin}= 1 - f_{rMS,bin}$.
The simulated distributions are compared with the observed ones by using a chi-squared test.
To derive the fraction of bMS and rMS binaries that better reproduce the observed distributions, we searched for the minimum of:
\begin{equation}
    \chi^2_{\rho}=\frac{\Sigma (\rho_{obs}-\rho_{syn})^2}{n_{bins}}
\end{equation}
where $n_{bins}$ is the number of $\Delta_{\rm F336W,F814W}$ intervals used to derive each distribution.

\begin{figure*}
    \centering
     \includegraphics[width=1.\linewidth]{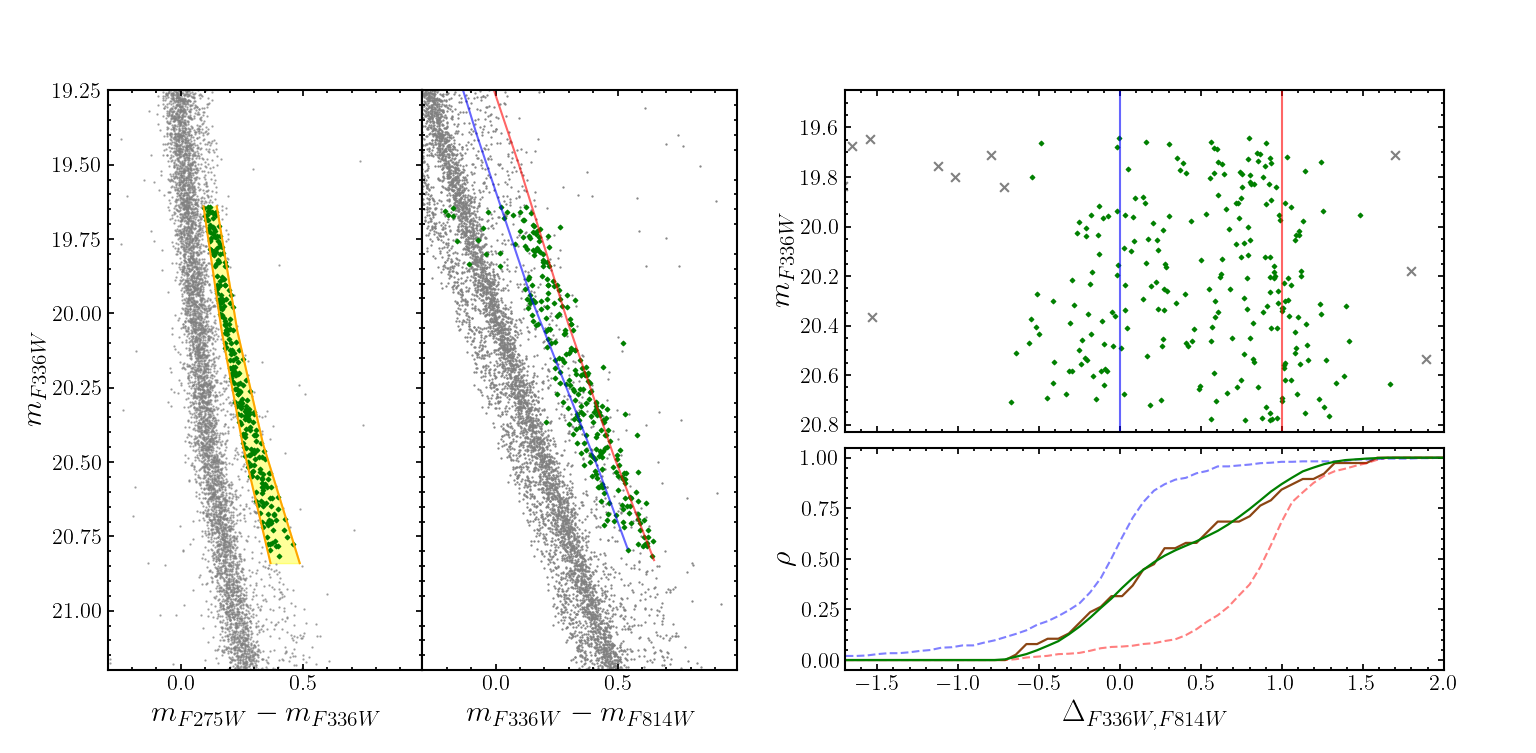}
     \caption{Simulated $m_{\rm F336W} vs.\,m_{\rm F275W}-m_{\rm F336W}$ (left) and $m_{\rm F336W} vs.\,m_{\rm F336W}-m_{\rm F814W}$ CMDs (middle) of NGC\,1850 that better reproduce the $\Delta_{F336W,F814W}$ cumulative distribution of the observed binaries. The verticalized $m_{\rm F336W}$ vs.\,$\Delta_{F336W,F814W}$ diagram is plotted on the top-right panel. The probable binaries, selected from the left-panel CMD, are colored green. Similarly to Fig.\,\ref{ngc1850_oss}, the orange lines in the left-panel CMD mark the boundaries of the yellow area, whereas the fiducial lines of the equal-luminosity bMS and rMS binaries are indicated with continuous blue and red lines, respectively. The bottom-right panel compares the $\Delta_{\rm F336W,F814W}$ cumulative distribution of the observed binaries (brown line), with the corresponding distribution for the simulated binaries. For completeness, we use the blue-dashed and the red-dashed lines to show the cumulative distributions derived from the simulated CMDs hosting bMS and rMS binaries alone, respectively. }
     \label{ngc1850_bestfit}
\end{figure*}

The minimum $\chi^2$ value inferred from the cumulative distribution provides a fraction of bMS binaries of $0.56\pm 0.06$, whereas the fraction of rMS binaries corresponds to $0.44\pm0.06$.
The uncertainties are derived as follows. We used the ASs to generate stars in the $m_{\rm F336W}$ vs.\,$m_{\rm F275W}-m_{\rm F336W}$ and $m_{\rm F336W}$ vs.\,$m_{\rm F336W}-m_{\rm F814W}$ CMDs. 
We simulated both single and binary stars by assuming that the 60\% and 40\% of the binaries belong to the bMS and the rMS, respectively.
We assumed a flat distribution of mass ratio for binaries and we used the fractions of rMS and bMS stars derived by \cite{milone2018}.

We generated 1,000 pairs of CMDs and, for each of them, we derived the fractions of bMS and rMS stars by following the same procedure adopted for real stars.
The uncertainties that we associated with the observed binary fractions are provided by the 1,000 determinations of binary fractions from the simulations.

Figure\,\ref{ngc1850_bestfit} illustrates the results corresponding to the best-fit simulation. As for the observed CMDs of Fig.\,\ref{ngc1850_oss}, we show the simulated $m_{\rm F336W}$ vs.\,$m_{\rm F275W}-m_{\rm F336W}$ and $m_{\rm F336W}$ vs.\,$m_{\rm F336W}-m_{\rm F814W}$ CMDs (left and middle panels) along with the $m_{\rm F336W}$ vs.\,$\Delta_{F336W,F814W}$ diagram for the selected binaries (top-right panel). The latter are selected from the left-panel CMD in close analogy with what we have done for real stars. The bottom-right panel of Fig.\,\ref{ngc1850_bestfit} compares the cumulative $\Delta_{F336W,F814W}$ distribution of the observed binaries with the corresponding distribution for the simulated binaries plotted in the top-right panel. 
For completeness, we show the cumulative distributions obtained from simulated CMDs hosting bMS binaries alone (blue dashed line) and rMS stars alone (red dashed line).

We notice that the fiducial lines of the red and blue MS do not coincide in the $m_{\rm F336W}$ versus $m_{\rm F275W}-m_{\rm F336W}$ CMD plotted in Fig.\,\ref{ngc1850_fids}. As a consequence, even by assuming that bMS and rMS binaries share the same mass-ratio distributions, the CMD region that we used to select the binaries (yellow dashed area of Fig.\,\ref{ngc1850_oss}) may include different fractions of bMS and rMS binaries relative to the total numbers of bMS and rMS stars.

Specifically, the fraction of rMS binaries would be a factor of 1.16 greater than that of bMS stars.
When we correct for this observational bias, we obtain that the fraction of bMS binaries in NGC\,1850 is $0.60\pm 0.06$, whereas the fraction of rMS binaries corresponds to $0.40 \pm 0.06$.

We applied the same method to the other investigated targets and listed the results in Table\,\ref{tab2}. We obtained fractions of bMS binaries of $0.59\pm0.10$ and $0.53\pm0.12$ for NGC\,1818 and NGC\,2164, which are similar to those observed in NGC\,1850.

These results, summarized in Table\,\ref{tab2}, rely on the  $\Delta_{\rm F336W,F814W}$ cumulative distribution of the selected binaries. 
For completeness, we derived the kernel-density distribution of the observed binaries by assuming a  Gaussian kernel with a dispersion of 0.2 and used it to estimate the fractions of bMS and rMS binaries. To do this, we compared the observed kernel-density distribution with the corresponding distributions of simulated binaries, in close analogy with what we did for the cumulative distribution.

\begin{figure}
    \centering
     \includegraphics[width=1.\linewidth]{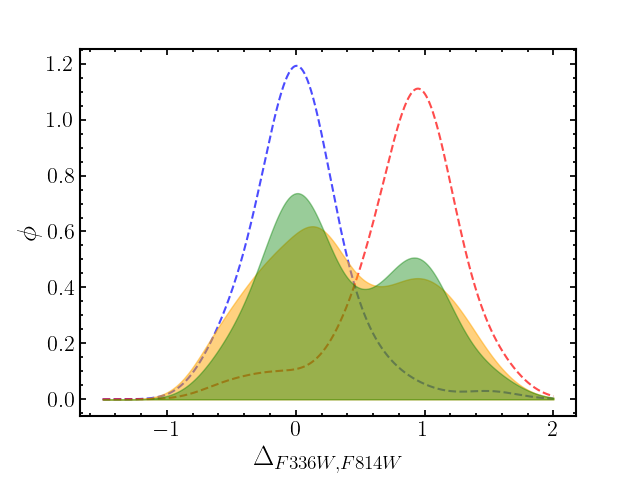}
     \caption{Comparison between the observed $\Delta_{\rm F336W,F814W}$ kernel distribution of the NGC\,1850 binaries (yellow-shaded area), and the corresponding distribution derived from the best-fitting simulated CMD (green-shaded area). The blue and red dashed lines indicate the distributions of binaries in a simulated CMD hosting bMS and rMS binaries alone, respectively. }
     \label{ngc1850_kde}
\end{figure}

We show in Fig.\ref{ngc1850_kde} the $\Delta_{\rm F336W,F814W}$ kernel distribution of the observed binaries in NGC\,1850  (shaded yellow area), and the corresponding distribution obtained from the simulated binaries that provides the best match with the observations (green shaded area). 
The dashed-blue and dashed-red distributions are derived from simulated CMDs that host bMS and rMS binaries alone. 

Noticeably, the observed distributions exhibit bimodal distributions with the two peaks corresponding to those of the distributions of bMS and rMS binaries.
The results coming from kernel analysis are in agreement with the ones derived using cumulative distribution. In particular, we obtained a fraction of bMS binaries of $0.60 \pm 0.06$, $0.50 \pm 0.10$,  and $0.53 \pm 0.13$ for NGC\,1850, NGC\,1818, NGC\,2164 respectively.

\begin{table}
    \centering
    \caption{Fractions of bMS and rMS stars among the overall binary population.}
    \begin{tabular}{ccc}
    \hline
    \hline
         Target& $f_{bMS,bin}$&  $f_{rMS,bin}$\\
         \hline
         NGC\,1818&   $0.59 \pm 0.10$ & $0.41\pm 0.10$  \\
         NGC\,1850&   $0.60 \pm 0.06$ & $0.40\pm 0.06$   \\
         NGC\,2164&   $0.53 \pm 0.12$ & $0.47\pm 0.12$  \\
         \hline
    \end{tabular}    
    \label{tab2}
\end{table}

\subsection{The fractions of red-MS and blue-MS stars} \label{sub:MSs}
To infer the fraction of rMS, bMS stars, and the fraction of binaries in the cluster field, we adopted the method described by \cite{milone2012b}.
 
\begin{figure}
    \centering
    \includegraphics[width=1.\linewidth]{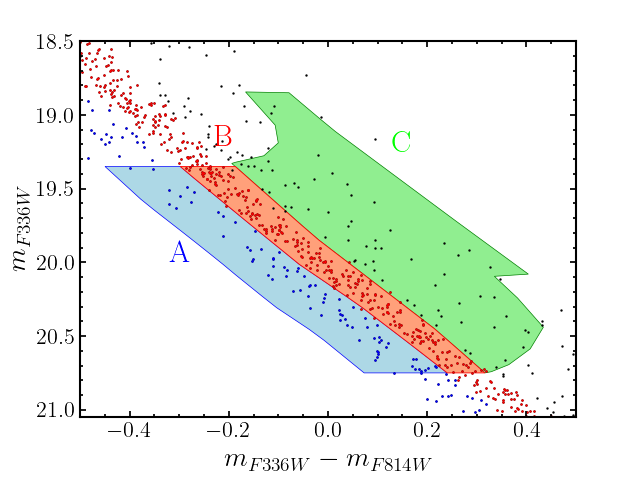}
    \caption{$m_{\rm F336W}$ vs.\,$m_{\rm F336W}-m_{\rm F814W}$ CMD of the cluster field of NGC\,1850. Blue, red, and green zones are delineated to encompass the majority of bMS, rMS, and binary stars, respectively.  }
    \label{ngc1818_frac}
\end{figure}

We defined three regions in the CMD, namely A, B, and C, represented by the blue, red, and green colors in Fig.\,\ref{ngc1818_frac}.

Regions A and B are mostly populated by blue MS stars and red MS stars with $19.4 < m_{F336W} < 20.75$ mag, respectively.
The limits of region A have been traced arbitrarily with the criteria of including most of the blue MS stars.
The red boundary of Region B has been evaluated starting from the rMS fiducial line and corresponds to the sequence of binaries with a mass ratio equal to 0.4. 
Region C includes binary systems with a mass ratio larger than 0.4. The red boundary of region C corresponds to the sequence of equal-mass rMS binaries shifted to red by the color error. 

Due to observational errors, each region is affected by contamination of stars from the other populations. Therefore, the total number of stars, N$_{\rm i}$ that we observe in each region can be indicated as
\begin{equation} \label{eq:1}
N_{\rm i=A,B,C}=N_{\rm bMS}f_{\rm bMS,i}+N_{\rm rMS}f_{\rm rMS,i}+N_{\rm bin}f_{\rm bin,i}
\end{equation}

Here, $f_{\rm bMS,i}$, $f_{\rm rMS,i}$, and $f_{\rm bin,i}$ are the fractions of bMS stars, rMS stars, and binaries in each region. These quantities are obtained from simulated CMDs composed of bMS, rMS, and binaries only, and derived by using ASs. The simulated binaries comprise the same fractions of bMS and rMS binaries that we inferred from the observations. 

The quantities $N_{\rm bMS}$, $N_{\rm rMS}$, and $N_{\rm bin}$ represent the total numbers of bMS stars, rMS stars, and binary systems, respectively.
To derive these values, we iteratively solved the system of linear equations (Eq.\,\eqref{eq:1}).
 At the first iteration, we assumed that the cluster hosts no binaries, whereas in the subsequent iterations, we adopted $N_{\rm bMS}$, $N_{\rm rMS}$, and $N_{\rm bin}$ as guesses to improve these values.
 Since the fractions of binaries in the regions A, B, and C of the CMD depend on the numbers of bMS and rMS stars, at each iteration we derived an improved determination of the quantity $f_{\rm bin,i}$. We iterated the procedure until the values of $N_{\rm bMS}$, $N_{\rm rMS}$, and $N_{\rm bin}$ obtained in two subsequent iterations differ by less than 0.001. 
The fractions of bMS and rMS stars and the fraction of binaries are listed in Table\,\ref{tab3} and confirm previous findings that the majority of MS stars belong to the rMS.

\subsection{The fraction of binaries among blue-MS and red-MS stars}
\label{sub:results}

\begin{figure}
    \centering
    \includegraphics[width=1.\linewidth]{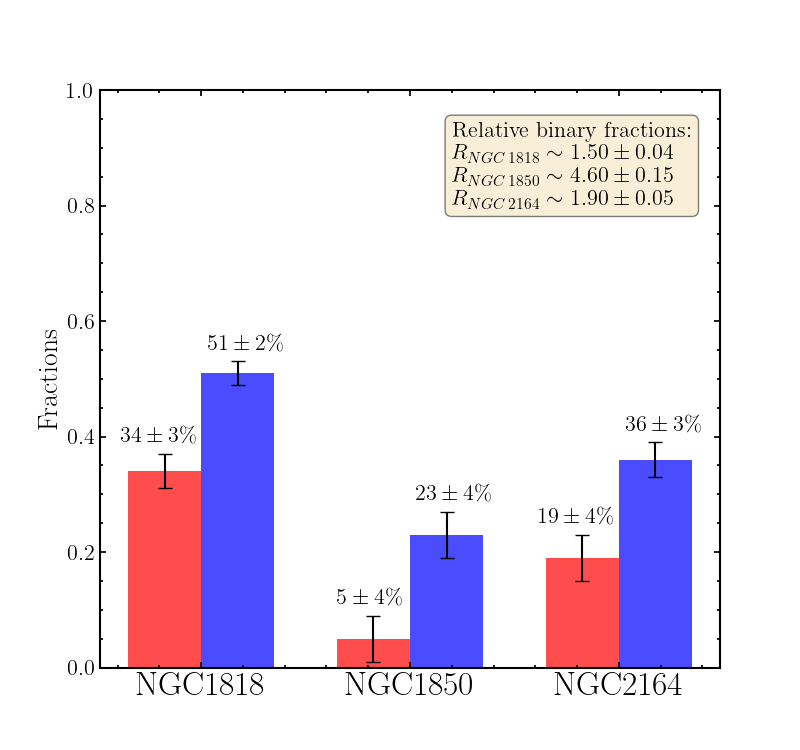}
    \caption{ Comparison between the fraction of binaries in bMS and rMS stars (blue and red bars respectively) in the studied clusters. The inset indicates the fraction of bMS stars relative to the rMS binary fraction, $F^{bin}_{bMS}/F^{bin}_{rMS}$ }
    \label{hist}
\end{figure}

 To derive the fractions of binaries among bMS and rMS stars, $F^{bin}_{bMS}$ and $F^{bin}_{rMS}$, we combined the results from the Section\,\ref{sub:BinMSs} and \ref{sub:MSs}.
 Specifically, we used the relation:
\vspace{-0.1 cm}
\begin{multline}
    F^{bin}_{bMS}= \frac{f_{bMS,bin}\cdot N_{bin} }{f_{bMS,bin}\cdot N_{bin} + N_{bMS}}\\
    F^{bin}_{rMS}= \frac{f_{rMS,bin}\cdot N_{bin} }{f_{rMS,bin}\cdot N_{bin} + N_{rMS}}\\
\end{multline}
\vspace{-0.1 cm}
In NGC\,1850, we obtained a fraction of binaries among the bMS and rMS stars of $23\pm4\%$ and $5\pm4\%$, respectively.
The results, presented in Table\,\ref{tab3} and illustrated in Figure\,\ref{hist}, reveal a predominance of bMS binaries in all clusters. Specifically, the ratio between $F^{bin}_{bMS}$  and $F^{bin}_{rMS}$ ranges from $\sim 4.60 \pm 0.15$ in NGC\,1850 to $\sim1.50\pm 0.04$ and $\sim1.90\pm 0.05  $ for NGC\,1818 and NGC\,2164, respectively.

\begin{table*}
    \centering
    \caption{
     Fractions of bMS stars, rMS stars, and binaries in the studied clusters.      }
    {\renewcommand{\arraystretch}{1.5}
    \begin{tabular}{cccccccc} 
        \hline
        \hline
         Target&  $F_{bMS}$ & $F_{rMS}$ & $F_{bin}$&  $F^{bin}_{bMS}$ &$F^{bin}_{rMS}$ &  Difference  & p-value\\ 
         & & & & & & $(F^{bin}_{bMS}-F^{bin}_{rMS})$ & \\
         \hline
         NGC\,1818&  $0.24 \pm 0.01$ & $0.34 \pm 0.02$ &  $0.42 \pm 0.02$ & $0.51 \pm 0.02$  & $0.34 \pm 0.03$ & $0.17 \pm 0.04$ & $<0.001$ \\ 
         NGC\,1850&  $0.20 \pm 0.01$ & $0.70 \pm 0.03$ &  $0.10 \pm 0.04$ & $0.23 \pm 0.04$  & $0.05 \pm 0.04$ & $0.18 \pm 0.06$ & $<0.001$ \\ 
         NGC\,2164&  $0.23 \pm 0.01$ & $0.52 \pm 0.02$ &  $0.25 \pm 0.03$ & $0.36 \pm 0.03$  & $0.19 \pm 0.04$ & $0.17 \pm 0.05$ & $<0.001$\\ 
         \hline
    \end{tabular}}
    %\vspace{0.2cm}
    \tablefoot{The last columns list the fraction of binaries among bMS and rMS stars, their difference, and their p-values based on the t-test.}
    \label{tab3}
\end{table*}

\section{Summary and conclusions} \label{sec:summary}

It is now widely accepted that star clusters younger than $\sim$600 Myr exhibit split MSs. The rMSs, which include the majority of MS stars, are populated by fast-rotating stars, while the bMSs consist of slow-rotating stars \cite[e.g.][and references therein]{milone2022, li2024}. Despite this, the origin of stars with varying rotation rates remains poorly understood. The main mechanisms proposed to explain the split MS include interactions in binary systems, the evolution of pre-MS stars, and stellar mergers. In this context, binary systems composed of bMS and rMS stars may provide insights to disentangle the different formation scenarios. Indeed, each scenario predicts different patterns of binarity between the two main sequences. Specifically, a predominance of binaries among bMS stars would support the possibility that interactions in binary star systems are responsible for stellar braking \citep{dantona2015, dantona2017}, whereas if the split MS is due to the early evolution of pre-MS stars we would expect similar fractions of bMS and rMS binaries \citep{bastian2020, kamann2021}. In contrast, in the scenario proposed by \citet{wang2022}, where the merging of binary systems is responsible for the origin of bMS stars, we would expect a predominance of binaries among rMS stars.

In this work, we present the first photometric derivation of the frequency of binaries among the bMS and rMSs of three young LMC clusters, namely NGC\,1818, NGC\,1850, and NGC\,2164.
To do this, we used high-precision photometry in the F814W, F336W, and F225W (or F275W) bands of UVIS/WFC3 of the {\it HST} \citep{milone2023}  refining the approach initially developed by \cite{milone2020} for investigating binary stars among multiple populations in Galactic GCs.  

In a nutshell, we utilized the $m_{\rm F336W}$ vs.\,$m_{\rm F225W}-m_{\rm F336W}$ (or $m_{\rm F275W}-m_{\rm F336W}$) CMDs, where the two MSs nearly overlap, to identify a sample of binaries composed of stars with similar luminosities.
We then analyzed these binaries in the selected binaries in the $m_{\rm F336W}$ vs.\,$m_{\rm F336W}-m_{\rm F814W}$ CMDs, which maximize the color separation between bMS and rMS stars. 
We estimated the fractions of binaries among bMS and rMS stars by comparing the $m_{\rm F336W}-m_{\rm F814W}$  color distribution of the observed binaries with grids of simulated CMDs that account for various binary fractions among bMS and rMS stars. 

We observe a predominance of binaries among the bMS stars in all the studied clusters. The fractions of bMS binaries exceed those of rMS binaries by $\sim$1.5 times in NGC\,1818, 4.6 times in NGC\,1850, and 1.9 times in NGC\,2164.
Noticeably, these results are qualitatively similar to what is observed in the Galactic field, where the binaries with periods between 4 and 500 days rotate much slower than the single stars \cite{abt2004}.

To understand the origin of the split MS in young Magellanic Cloud star clusters, \cite{dantona2015} and \cite{dantona2017} proposed that all cluster stars are born as fast rotators.
Interactions in binary systems may be responsible for stellar braking, with tidal interactions causing braking effects that begin in the core and extend outward. When this braking process affects the stellar envelope and photosphere, the stars reach their final position on the bMS.
Our finding of a predominance of binaries among the bMS stars aligns with the idea proposed by D'Antona and colleagues, suggesting that their scenario represents the primary pathway for the formation of bMS stars.

Recently, \cite{kamann2021} exploited multi-epoch observations of NGC\,1850 collected with the MUSE integral
field spectrograph at the Very Large Telescope to investigate the role of binarity in the origin of the split MS \citep[see also][]{saracino2023}.
They concluded that the slow and rapidly rotating populations host similar fractions of binaries of $\sim 5\%$, in tension with the outcomes of our photometric study ons NGC\,1850 and with the results by \cite{abt2004} on the Galactic field.

However, we observe that Kamann and collaborators' spectroscopic analysis primarily detects hard binaries. In contrast, our study's sample of photometric binaries encompasses all binaries with large mass ratios, regardless of their periods and semimajor axes.
 One possible way to reconcile the results of \cite{kamann2021} with our findings is by considering that the soft binaries mostly populate the bMS of NGC\,1850. In this scenario, the formation of bMS stars could be linked to tidal interactions within soft binaries, as proposed by \cite{He2023} and \cite{yang2021}.
 
 Finally, we notice that the observed predominance of binaries may challenge the possibility that stellar mergers of MS stars are the main responsible for the formation of bMS stars \citep{wang2022}. Indeed, in this case, we could expect a lack of binaries among the bMS.
 However, the lack of a quantitative analysis on the role of triple stellar systems in the merging process prevents us from firm conclusions.

\begin{acknowledgements}
We thank the anonymous referee for various suggestions that improved the quality of the manuscript.
This work has been funded by the European Union – NextGenerationEU RRF M4C2 1.1 (PRIN 2022 2022MMEB9W: "Understanding the formation of globular clusters with their multiple stellar generations", CUP C53D23001200006), 
from INAF Research GTO-Grant Normal RSN2-1.05.12.05.10 -  (ref. Anna F. Marino) of the "Bando INAF per il Finanziamento della Ricerca Fondamentale 2022", and from the European Union’s Horizon 2020 research and innovation programme under the Marie Skłodowska-Curie Grant Agreement No. 101034319 and from the European Union – NextGenerationEU (beneficiary: T. Ziliotto).
We thank the anonymous referee 
\end{acknowledgements}

\bibliographystyle{aa}
\bibliography{mybib}{}

\end{document}